\documentclass[10pt,letterpaper]{article}
\usepackage{subcaption}
\usepackage{opex3}
\usepackage{amsmath}
\usepackage{amssymb}
\usepackage{cite}
\usepackage{color}
\usepackage{array}
\usepackage{upgreek}

\newcommand{\mb}[1]{\mathbf{#1}}

\usepackage{tikz}
\usetikzlibrary{patterns}
\usepackage{booktabs}
\begin{document}

%% START HERE
%%%%%%%%%%%%%%%%%% title page information %%%%%%%%%%%%%%%%%%
\title{Effects of Raman scattering and attenuation in silica fiber-based parametric frequency conversion}

\author{S\o{}ren M. M. Friis,$^{*}$ Lasse Mejling, and Karsten Rottwitt}

\address{Department of Photonics Engineering, Technical University of Denmark, \\ 2800 Kongens Lyngby, Denmark}

\email{$^{*}$smmf@fotonik.dtu.dk} %% email address is required



\begin{abstract*}
Four-wave mixing in the form of Bragg scattering (BS) has been predicted to enable quantum noise less frequency conversion by analytic quantum approaches. Using a semi-classical description of quantum noise that accounts for loss and stimulated and spontaneous Raman scattering, which are not currently described in existing quantum approaches, we quantify the impacts of these effects on the conversion efficiency and on the quantum noise properties of BS in terms of an induced noise figure (NF). We give an approximate closed-form expression for the BS conversion efficiency that includes loss and stimulated Raman scattering, and we derive explicit expressions for the Raman-induced NF from the semi-classical approach used here. \vspace{1cm}
\end{abstract*}

%\ocis{(190.4380) Nonlinear optics, four-wave mixing; (270.2500) Fluctuations, relaxations, and noise.}

%%%%%%%%%%%%%%%%%%%%%%% References %%%%%%%%%%%%%%%%%%%%%%%%%

%%%%%%%%%%%%%%%%%%%%%%%%%%  body  %%%%%%%%%%%%%%%%%%%%%%%%%%
\section{Introduction}
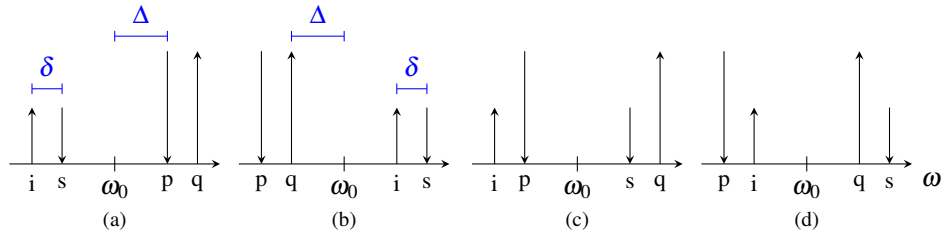
\begin{figure}[!b]
\centering
\begin{tikzpicture}
% Build axes
\draw [-stealth] (-6,0)--(-3.2,0);
	\draw (-4.6,0.1)--(-4.6,-0.1) node [below] {$\omega_0$};
	\node at (-4.6,-0.75) {\footnotesize (a)};
\draw [-stealth] (-2.95,0)--(-0.15,0);
	\draw (-1.55,0.1)--(-1.55,-0.1) node [below] {$\omega_0$};
	\node at (-1.55,-0.75) {\footnotesize (b)};	
\draw [-stealth] (3.2,0)--(6,0) node [below right] {$\omega$};
	\draw (4.6,0.1)--(4.6,-0.1) node [below] {$\omega_0$};
	\node at (4.6,-0.75) {\footnotesize (d)};
\draw [stealth-] (2.95,0)--(0.15,0);
	\draw (1.55,0.1)--(1.55,-0.1) node [below] {$\omega_0$};	
	\node at (1.55,-0.75) {\footnotesize (c)};	
	
% Build waves
\draw [-stealth] (-5.7,0)--(-5.7,0.75);
	\node at (-5.7,-0.25) {\small i};
\draw [stealth-] (-5.3,0)--(-5.3,0.75);
	\node at (-5.3,-0.25) {\small s};
\draw [stealth-] (-3.9,0)--(-3.9,1.5);
	\node at (-3.9,-0.25) {\small p};
\draw [-stealth] (-3.5,0)--(-3.5,1.5);
	\node at (-3.5,-0.25) {\small q};

\draw [stealth-] (-2.65,0)--(-2.65,1.5);
	\node at (-2.65,-0.25) {\small p};
\draw [-stealth] (-2.25,0)--(-2.25,1.5);
	\node at (-2.25,-0.25) {\small q};
\draw [-stealth] (-0.85,0)--(-0.85,0.75);
	\node at (-0.85,-0.25) {\small i};
\draw [stealth-] (-0.45,0)--(-0.45,0.75);
	\node at (-0.45,-0.25) {\small s};

\draw [stealth-] (5.7,0)--(5.7,0.75);
	\node at (5.7,-0.25) {\small s};
\draw [-stealth] (5.3,0)--(5.3,1.5);
	\node at (5.3,-0.25) {\small q};
\draw [-stealth] (3.9,0)--(3.9,0.75);
	\node at (3.9,-0.25) {\small i};
\draw [stealth-] (3.5,0)--(3.5,1.5);
	\node at (3.5,-0.25) {\small p};

\draw [-stealth] (2.65,0)--(2.65,1.5);
	\node at (2.65,-0.25) {\small q};
\draw [stealth-] (2.25,0)--(2.25,0.75);
	\node at (2.25,-0.25) {\small s};
\draw [stealth-] (0.85,0)--(0.85,1.5);
	\node at (0.85,-0.25) {\small p};
\draw [-stealth] (0.45,0)--(0.45,0.75);
	\node at (0.45,-0.25) {\small i};

% Accessories shift={(3 cm,5 cm)}
\draw [color =blue] (-5.7,1)--(-5.3,1);
\draw [color =blue] (-5.7,1.07)--(-5.7,0.93);
\draw [color =blue] (-5.3,1.07)--(-5.3,0.93);
	\node [color =blue] at (-5.5,1.3) {$\delta$};

\draw [color =blue,shift={(4.85,0)}] (-5.7,1)--(-5.3,1);
\draw [color =blue,shift={(4.85,0)}] (-5.7,1.07)--(-5.7,0.93);
\draw [color =blue,shift={(4.85,0)}] (-5.3,1.07)--(-5.3,0.93);
	\node [color =blue,shift={(4.85,0)}] at (-5.5,1.3) {$\delta$};

\draw [color =blue,shift={(0,0.7)}] (-2.25,1)--(-1.55,1);
\draw [color =blue,shift={(0,0.7)}] (-2.25,1.07)--(-2.25,0.93);
\draw [color =blue,shift={(0,0.7)}] (-1.55,1.07)--(-1.55,0.93);
	\node [color =blue] at (-1.9,2) {$\Delta$};
	
\draw [color =blue,shift={(-2.35,0.7)}] (-2.25,1)--(-1.55,1);
\draw [color =blue,shift={(-2.35,0.7)}] (-2.25,1.07)--(-2.25,0.93);
\draw [color =blue,shift={(-2.35,0.7)}] (-1.55,1.07)--(-1.55,0.93);
	\node [color =blue,shift={(-2.35,0)}] at (-1.9,2) {$\Delta$};
\end{tikzpicture}
\caption{Sketches of selected frequency configurations performing down-conversion from signal (s) to idler (i); the two pumps, p and q, need not have equal magnitudes. $\delta$ is the frequency separation between the two side-bands in either of (a)--(d), and $\Delta$ is the frequency separation between the zero-dispersion frequency and the closest wave component on either side. By changing $\omega_{\rm s} \leftrightarrow \omega_{i}$ and $\omega_{p} \leftrightarrow\omega_{q}$, up-conversion is achieved} \label{fig_1}
\end{figure}

Parametric processes offer a range of important potential applications in future all-optical communication systems: low-noise phase-insensitive and phase-sensitive amplification \cite{Tong_2010a,Tong_2011a}; the ability to regenerate noisy optical signals \cite{Croussore_2008a,Skold_2008a}; and alteration of the temporal mode profile as well as the frequency of optical quantum states, while preserving other properties of the quantum state of light \cite{Tanzilli,Raymer_2012a}. The third-order nonlinearity of optical fibers has been shown to enable tunable frequency conversion by four-wave mixing (FWM) in a special configuration known as Bragg scattering (BS) \cite{Inoue_1994a}. In BS, one uses two pumps p and q, which interact with a signal s and an idler i fulfilling energy conservation $\omega_{\rm p}+\omega_{\rm s}=\omega_{\rm q}+\omega_{\rm i}$, where $\omega_j$ is the angular frequency of the electric field for $j = \rm \{p,q,s,i\}$, see Fig. \ref{fig_1}. Even though BS is driven by two strong pumps, the process has been predicted by analytic quantum approaches to be free of additional quantum noise \cite{McK_2005a,McK_2005b,Mcguinness_2010a,Mejling_2012a} as opposed to phase-insensitive parametric amplifiers or Raman and Erbium-doped fiber amplifiers, where amplified spontaneous emission induce a 3-dB noise figure (NF) in the high gain limit. These predictions are based on FWM being effectively phase-matched and the only influence on the wave components. However, material and waveguide dispersion of the third-order nonlinear device in which BS is realized may induce a significant phase mismatch and thereby reduce the conversion to take place in a confined frequency window. Also, amorphous materials such as silica and chalcogenide ($\rm As_2S_3$) have broad Raman response spectra that typically extend beyond this limit of phase matching \cite{Clark_2012}. Finally, attenuation in the nonlinear medium reduces the efficiency of BS and therefore limits the achievable output. 

BS has been studied extensively for both classical \cite{Gnauck_2006,Mechin2006} and quantum signals \cite{Mcguinness_2010a,McK_2012a,Clark_2013a}, and the effects of stimulated and spontaneous Raman scattering (SRS and SpRS) have been studied in parametric amplifiers \cite{Tong_2010a,Voss_2004a,Voss_2006a,Friis_2013} and for photon-pair generation by degenerate four-wave mixing \cite{Collins_2012a,Lin_2006}, but little attention has been given to their impacts on frequency conversion using BS, which is mainly due to the difficulty of implementing Raman scattering in quantum descriptions of FWM.

In this paper, we use a semi-classical approach \cite{Friis_2013} to produce realistic predictions for the conversion efficiency (CE) and noise properties of BS in silica-based highly nonlinear fibers. The semi-classical approach simulates coherent states by classical field ensembles that display correct Gaussian statistics. One ensemble for each wave component is then propagated through the classical equations describing FWM including dispersion, loss, and Raman scattering. During propagation in the fiber, fluctuations are added to the ensemble, which account for spontaneous emission, and thus distort the performance of the FWM process. Note that such semi-classical approaches can predict the quantum noise properties of classical signals but they do not capture the quantum nature of single-photon states.

In Fig. \ref{fig_1}, four possible frequency configurations of BS are shown: in (a) and (b), the signal and idler are placed on the Stokes (S) and anti-Stokes (aS) side of the two pumps, respectively; (c) and (d) offer alternative configurations that have the same FWM properties but different Raman interaction. For all four setups, $\delta$ is the frequency separation between the two components of each side-bands, and $\Delta$ is the separation between the zero-dispersion frequency, $\omega_0$, and the closest wave component on either side of $\omega_0$. It is implied that the waves are placed symmetrically around $\omega_0$ to obtain phase matching independent on the dispersion slope of the fiber. In this work, we investigate only the two setups (a) and (b).

This paper is structured as follows: in Sec. 2, the classical propagation equations that describe FWM, SRS and loss in a fiber are derived and the approach for semi-classical modeling of quantum noise is outlined. In Sec. 3, we investigate the effects of fiber attenuation using realistic physical parameters. Finally in Sec. 4, we quantify the combined effects of FWM, SRS, and SpRS, and we show that Raman scattering has a significant impact on both the CE and noise properties of BS in a typical highly nonlinear silica fiber.

%%%%%%%%%%%%%%%%%%%%%%%%%%%
%
%
% SEC 2: Theory
%
%
%%%%%%%%%%%%%%%%%%%%%%%%%%%
\section{Theory}

\subsection{Propagation equations of FWM and Raman scattering}
Equations that describe the propagation of an electromagnetic field through nonlinear optical fibers including the effects of both FWM and Raman scattering can be derived directly from Maxwell's equations; using the approach of \cite{Kolesik2004}, a general propagation equation in a single mode nonlinear fiber for the amplitude of a continuous-wave electric field, $E_n$, at frequency $\omega_n$ is \cite{Poletti2008} 
\begin{align} \label{eq_1}
\frac{\partial E_n(z)}{\partial z} = \frac{i \omega_n{\rm e}^{ - i\beta_n z}}{4N_n} \iint {{\mb F}_n}^*(x,y) \cdot \mb{P}_n^{(3)}(\mb{r})\, dxdy,
\end{align}
where $\beta_n$ is the propagation constant at frequency $\omega_n$, and ${\mb r} = (x,y,z)$ is the position vector where $z$ is the longitudinal coordinate in the fiber and $x$ and $y$ are the transverse coordinates. In the derivation of Eq. \eqref{eq_1}, the total electric field was expanded in a set of continuous-wave components
\begin{align}
\mb{E}(\mb{r},t) = \frac{1}{2} \sum_{m} \frac{\mb{F}_m(x,y)}{N_m}E_n(z,t){\rm e}^{i\beta_m z -i\omega_m t} + {\rm c.c.},
\end{align}
where $\mb{F}_m(x,y)$ is a three-component vector holding the transverse field distribution functions of the three spatial components of the electric field, and $N_m$ is a normalization constant defined as \cite{Poletti2008}
\begin{align}
\frac{1}{4}\iint \left[ {\mb{F}_m}^* \times {\mb{H}_m} + {\mb{F}_m} \times {\mb{H}_m}^* \right ]\cdot \hat{\mb z}\, dxdy= {N_m}^2
\end{align} 
where $\mb{H}_n$ is the same for the magnetic field as $\mb{F}_n(x,y)$ is for the electric field. For a linearly polarized field, $\mb{F}_m(x,y) = F_m(x,y)\hat{\mb x}$, where $\hat{\mb x}$ is a unit vector pointing in the transverse direction x in the fiber, $N_m^2$ is calculated to be
\begin{align}\label{eq_104}
N_m^2 = \frac{c\epsilon_0 n_m^{\rm eff}}{2}\iint F(x,y)^2 dxdy,
\end{align}
where $n_m^{\rm eff}$ is the effective refractive index at frequency $\omega_m$. The time-independent third order nonlinear induced polarization $\mb{P}_n^{(3)}$ in Eq. \eqref{eq_1} is defined in terms of the total third order nonlinear induced polarization as
\begin{align} \label{eq_103}
\mb{P}^{(3)}(\mb{r},t) = \frac{1}{2} \sum_m \mb{P}_m^{(3)}(\mb{r}) {\rm e}^{-i\omega_m t} + {\rm c.c.}.
\end{align}
The total nonlinear induced polarization is written as \cite{Agrawal}
\begin{align}\label{eq_102}
\mb{P}^{(3)}(\mb{r},t) = \epsilon_0 \iiint_{-\infty}^{\infty} \mb{R}^{(3)}&(t-\tau_1,t-\tau_2,t-\tau_3)\, \vdots \,  \mb{E}(\mb{r},\tau_1)\mb{E}(\mb{r},\tau_2) \mb{E}(\mb{r},\tau_3) d\tau_1d\tau_2d\tau_3,
\end{align}
where the vertical dots denote the third order tensor product. The third order response function including both FWM and Raman scattering is \cite{Poletti2008},
\begin{align} \label{eq_101}
\mb{R}^{(3)}(t_1,t_2,t_3) = \chi^{(3)}\left( [1-f_R]\delta(t_1) + \frac{3}{2}f_Rh_R(t_1) \right) \delta(t_1-t_2) \delta(t_3)
\end{align}
under the assumption that only one term of the third order susceptibility, $\chi^{(3)}_{xxxx} \equiv \chi^{(3)}$, is non-zero, where $\chi^{(3)}$ is the total third order susceptibility, $f_R = 0.18$ is the Raman fraction, and $h_R(t)$ is the Raman response function.

In this paper, we consider four frequency components that are connected through energy conservation $\omega_1 + \omega_4 = \omega_2 + \omega_3$ and they interact through both FWM and Raman scattering. Inserting Eq. \eqref{eq_101} into Eq. \eqref{eq_102}, collecting terms in Eq. \eqref{eq_102} that oscillate at $\omega_j$ where $j=\{1,2,3,4\}$ \cite{Headley1996}, then using Eq. \eqref{eq_103} to insert $\mb{P}_j^{(3)}$ into Eq. \eqref{eq_1}, and finally using Eq. \eqref{eq_104}, the propagation equation for the field amplitude at $\omega_1$ is derived,
\begin{align} \label{eq_105}
\begin{split}
\frac{\partial E_1}{\partial z} = - \frac{\alpha}{2}E_1 + \frac{i \omega_1n_2}{c A_{\rm eff}}& \left[ |E_1|^2 E_1 + (2-f_R) E_1\sum_{n=2}^4 |E_n|^2 \right. \\
 &\hspace{1cm} \left. + 2(1-f_R)E_2 E_3 E_4^* {\rm e}^{i\Delta \beta z} + f_RE_1\sum_{n=2}^4 \tilde{h}_R(\Omega_{n1})|E_n|^2\right]
\end{split}
\end{align} 
where $n_2 = 3\chi^{(3)}\epsilon_0/(4n_{\rm eff}^2 c)$ is the nonlinear refractive index, $A_{\rm eff}$ is the usual effective area of the nonlinear fiber \cite{Agrawal}, $\Delta \beta$ is the phase mismatch elaborated below (note that in the equations of $\omega_2$ and $\omega_3$, the sign in front of $\Delta \beta$ is opposite), and $\tilde{h}_R(\Omega_{nm})$ is the Fourier transform of $h_R(t)$ evaluated at $\Omega_{nm} = \omega_n - \omega_m$. The linear loss term with coefficient $\alpha$ has been added and it is assumed common for all frequencies.

Equation \eqref{eq_105} and the corresponding equations of $\omega_2$ through $\omega_4$ are applied to case (a) in Fig. \ref{fig_1} by setting $\omega_1 = \omega_{\rm i}$, $\omega_2 = \omega_{\rm s}$, $\omega_3 = \omega_{\rm p}$, and $\omega_4 = \omega_{\rm q}$, and similarly for case (b). The time-domain Raman response function can be approximated by a single damped oscillator function \cite{Blow_1989}, but it turns out that such a simple description is not accurate for SpRS close to the pumps. Therefore, we use an extended response function,
\begin{eqnarray} \label{eq_21}
h_{\rm R}(t) = \sum_{j = 1}^{13}\frac{b_j}{\omega_j}\exp(-\eta_j t)\exp(-\Gamma_j^2 t^2/4)\sin(\omega_j t)\Theta(t),
\end{eqnarray}
where the coefficients $b_j$, $\omega_j$, $\eta_j$, and $\Gamma_j$ are determined in \cite{Hollenbeck_2002} and valid for a silica-core fiber \cite{Stolen_1989}, and $\Theta(t)$ is the heaviside step function for which $\Theta(t) = 0$ for $t<0$ and $\Theta(t) = 1$ for $t \geq 0$. To implement Eq. \eqref{eq_21} in Eq. \eqref{eq_105}, the Fourier transform is applied and the resulting imaginary part is used while the real part is disregarded for simplicity. According to \cite{Hsieh_2007} a significant effect of the real part of the Raman susceptibility is only observed when using a strong pulsed pump; the comparatively weak continuous-wave pumps used in this paper justify neglecting the real part of the Raman susceptibility.

By expanding the wave-number of each wave component in a Taylor series around the zero-dispersion frequency, the phase-mismatch $\Delta \beta = \beta_2 + \beta_3 - \beta_1 - \beta_4$ is written in terms of $\delta$ and $\Delta$ from Fig. \ref{fig_1} as
\begin{eqnarray} \label{eq_22}
\Delta \beta \approx -\frac{\beta_4}{12}\delta (2\Delta + \delta)\left( 2\Delta^2+2\Delta \delta + \delta^2 \right).
\end{eqnarray}
Because of the symmetry of the BS configuration around the zero-dispersion frequency, all odd dispersion terms cancel, which means $\beta_4$ is the lowest one that is used in the calculations. If we consider only the FWM terms, it is instructive to express the CE in terms of the phase mismatch. We assume the two strong pumps to have constant amplitudes such that
\begin{eqnarray}\label{eq_3}
E_{\rm p}(z) &=& \sqrt{P_{\rm p}}\exp(i\gamma_p[P_{\rm p} + (2-f_R)P_{\rm q}]z),\\
E_{\rm q}(z) &=& \sqrt{P_{\rm q}}\exp(i\gamma_q[P_{\rm q} + (2-f_R)P_{\rm p}]z), \label{eq_4}
\end{eqnarray}
where $P_{\rm p}$ and $P_{\rm q}$ are the constant pump powers and $\gamma_j = n_2 \omega_j/(cA_{\rm eff})$. By solving the resulting equations for the signal and idler, neglecting loss and Raman scattering, one easily finds
\begin{eqnarray} \label{eq_23}
{\rm CE}(z) = \frac{|E_{\rm i}(z)|^2}{|E_{\rm s}(0)|^2} =  \frac{\eta_{\rm i}^2}{(\kappa/2)^2 + \eta_{\rm i}\eta_{\rm s}} \sin^2(gz)%\left(1-\left[\frac{\kappa}{2g}\right]^2\right)\sin^2(gz),
\end{eqnarray}
where $\eta_j = 2(1-f_R)\gamma_j \sqrt{P_{\rm p}P_{\rm q}}$ is the effective nonlinear strength, $g^2 = \eta_{\rm i}\eta_{\rm s} + (\kappa/2)^2$ is the phase-mismatched conversion coefficient, and 
\begin{align}
\kappa = \Delta \beta \pm (1-f_R)(\gamma_p + \gamma_q) (P_{\rm q}-  P_{\rm p}) \mp (1-f_R)(\gamma_{\rm i}P_{\rm q} - \gamma_{\rm s}P_{\rm p}), 
\end{align}
where the first `$+$' (second `$-$') corresponds to case (a) and (c) of Fig. \ref{fig_1}, and the first `$-$' (second `$+$') corresponds to (b) and (d). Maximum conversion from signal to idler is achieved for $\kappa = 0$, which gives $\rm{CE} = \eta_{\rm i}/\eta_{\rm s}$, which further is the same as full conversion in photon numbers. This condition can be met experimentally by adjusting the difference in pump powers to counter-balance the phase mismatch if the values of $\delta$ and $\Delta$ are not too large relative to the fourth-order dispersion. Alternatively, a broad bandwidth of phase matching can be achieved through dispersion engineering, as demonstrated recently in a dispersion shifted fiber \cite{Farsi_2015}, or special phase matching properties across small \cite{Friis_2016} or large \cite{Demas_2015} bandwidths can be achieved using higher order modes. If $\kappa \approx 0$ is valid, one may, with a few simplifying assumptions, derive an approximate analytic expression for the CE where Raman scattering and loss are included. In the pump equations, energy exchange terms, cross-phase modulation terms of the signal and idler, the effect of loss on the phase modulation terms of the pumps, and the Raman interaction between the pumps are disregarded. In the signal and idler equations, only the small phase modulation terms of the signal and idler are disregarded. The CE from signal to idler becomes (after some calculations)
\begin{align} \label{eq_14}
{\rm CE} = \frac{|E_{\rm i}(z)|^2}{|E_{\rm s}(0)|^2} &=\frac{\eta_{\rm i}^2}{\mu^2} \exp\left[(f_{\rm s}+f_{\rm i})z_{\rm eff}\right] \exp(-\alpha z) \sin^2\left( \mu z_{\rm eff} \right)\\ \label{eq_13}
&\approx \frac{\eta_{\rm i}^2}{\mu^2}\exp\left[(f_{\rm s}+f_{\rm i}-\alpha)z\right]\sin^2(\mu z),
\end{align}
where $\mu^2 = \eta_{\rm i}\eta_{\rm s}-(f_{\rm i} - f_{\rm s})^2/4 $ is the phase matched conversion coefficient, $z_{\rm eff} = (1-\exp[-\alpha z])/\alpha$ is the effective position in the fiber, and $f_{\rm i} = i\gamma_{\rm i}f_R(\tilde h_R(\Omega_{\rm pi})P_{\rm p} + \tilde h_R(\Omega_{\rm qi})P_{\rm q})$ and $f_{\rm s} = i\gamma_{\rm s}f_R(\tilde h_R(\Omega_{\rm ps})P_{\rm p} + \tilde h_R(\Omega_{\rm qs})P_{\rm q})$ are the Raman contributions to the signal and idler, respectively. In Eq. \eqref{eq_13}, it was assumed that $\alpha z \ll 1$, which is valid for typical lengths of highly nonlinear fibers.

\subsection{Spontaneous Raman scattering \label{sec_1}}
\begin{figure}
\centering
\includegraphics[width = 0.99\textwidth]{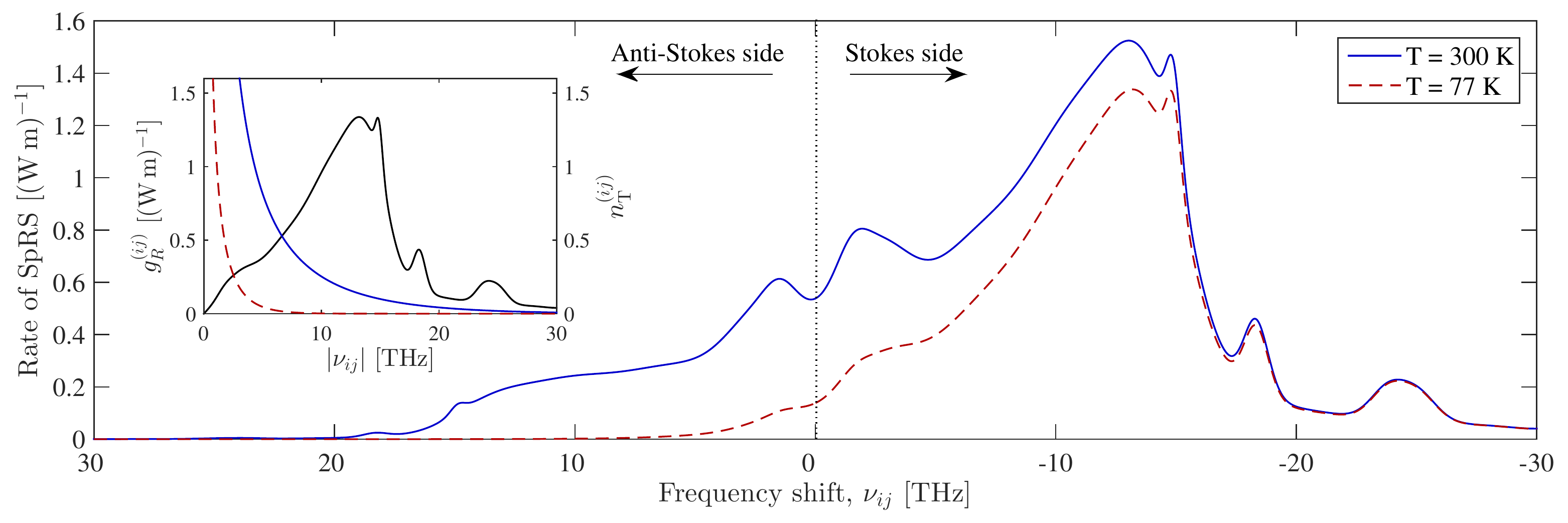}
\caption{Rate of SpRS versus real frequency shift $\nu_{ij} = \Omega_{nm}/2\pi$ from a pump p or q for temperatures $T = 300$ K (solid blue curve) and $T = 77$ K (dashed red curve). The inset shows the Raman gain coefficient $g^{(nm)}_{R}$ for silica core fibers (solid black curve) and the phonon equilibrium numbers $n^{(nm)}_{\rm T}$ for the same temperatures as in the main plot (with same line styles).}\label{fig_2}
\end{figure}

SpRS has been studied earlier \cite{Rottwitt_2003,Stolen_1970}, and a model for the accumulation of amplified spontaneous emission (ASE) seeded by the Raman process at frequency $\omega_m$ on the S side of a given pump at frequency $\omega_n$ of power $P_n$ is
\begin{eqnarray} \label{eq_31}
P_{\rm ASE,S}(z) = \hbar \omega_{\,m} B_0 (n^{(nm)}_{\rm T}+1)g^{(nm)}_{ R} P_n\, z,
\end{eqnarray}
where $B_0$ is the frequency bandwidth of the wave component at frequency $\omega_{\, m}$, $g^{(nm)}_{R} = 2 \gamma_{m} f_R \tilde h_R(\Omega_{nm})$ is the Raman gain coefficient, and $n^{(nm)}_{\rm T}(\Omega_{nm})=[\exp(\hbar |\Omega_{nm}|/k_{\rm B}T)-1]^{-1}$ is the phonon equilibrium number at a pump-signal frequency separation of $\Omega_{nm}$, where $\hbar$ is Plancks reduced constant, $k_{\rm B}$ is Boltzmanns constant, and $T$ is the temperature \cite{Kidorf_1999}. On the aS side, the corresponding expression is
\begin{eqnarray}
P_{\rm ASE,aS}(z) = \hbar \omega_{\, m} B_0 n^{nm}_{\rm T}|g^{(nm)}_{R}| P_n\,z.
\end{eqnarray}
The $(n^{(nm)}_{\rm T} +1)$-term in Eq. \eqref{eq_31} means that SpRS on the S side does not require any phonons present but on the aS side the $n^{(nm)}_{\rm T}$-term gives SpRS a proportional dependence on the number of phonons. We define the rates of SpRS for the S and aS processes as $(n^{(nm)}_{\rm T} +1)g^{(nm)}_{R}$ and $n^{(nm)}_{\rm T}|g^{(nm)}_{R}|$, respectively, and they are plotted in Fig. \ref{fig_2}, which shows these quantities at room (solid blue) and at liquid nitrogen (dashed red) temperature versus frequency shift $\Omega_{nm}$ from one of the pumps, p or q. In contrast to SRS, the rate of SpRS is asymmetric around the pump, which is important to consider when small signals are situated simultaneously on both sides of the pump, e.g. as they may be in FWM. Lowering the temperature reduces the rate of SpRS significantly on the aS side as expected, whereas on the S side a significant reduction occurs only close to the pump.

\subsection{Simulating quantum noise classically \label{sec_23}}
The quantum noise properties of parametric processes \cite{McK_2005b} can be predicted by semi-classical methods \cite{Gordon_1963}. One approach is to create field ensembles for simulating quantum coherent states at the input of the fiber and then propagate each element of the ensemble through classical equations (as derived in the previous section) in parallel; while propagating through the fiber, fluctuations must be added to the ensembles at every numerical step to account for SpRS and loss \cite{Friis_2013}. The inherent fluctuations of a quantum field is represented in the ensemble, and the signal-to-noise ratio (SNR) is evaluated as
\begin{eqnarray} \label{eq_36}
{\rm SNR} = \frac{\langle |A_{\rm ens}|^2 \rangle^2 }{{\rm Var} \left( |A_{\rm ens}|^2 \right)},
\end{eqnarray}
where $A_{\rm ens}$ is the complex valued classical field ensemble and Var($\cdot$) denotes the ensemble variance. The induced noise figure (NF) is then defined as $\rm NF = SNR_{in}/SNR_{out}$. The induced NF for the BS process is defined as $\rm SNR_{s,in}/SNR_{i,out}$, i.e. the SNR of the signal at the input relative to the SNR of the idler at the output. A coherent state is simulated classically by defining the field amplitude ensemble as
\begin{eqnarray} \label{eq_24}
A_{\rm ens} = x_0 + \delta x + i(p_0+\delta p),
\end{eqnarray}
where $\delta x$ and $\delta p$ are the quadrature fluctuation variables that follow Gaussian distributions with the properties $\langle \delta x \rangle=\langle \delta p \rangle = 0$ and $\langle \delta x^2 \rangle=\langle \delta p^2 \rangle = \hbar \omega B_0 /4$ \cite{Gerry}, and $\langle |A_{\rm ens}|^2 \rangle = x_0^2 + p_0^2 + \hbar \omega B_0/2$ is the mean field power. The variables $x_0$ and $p_0$ are the quadrature mean values, and the last term, $\hbar \omega B_0/2$, is the explicit inclusion of the vacuum state energy, which is an artifact of the semi-classical model. The magnitude, however, is so small that no significant error in the mean field power can be observed. To obtain reliable statistics, we use ensembles of $5\cdot 10^4$ elements in all simulations.

By propagating the ensemble of Eq. \eqref{eq_24} through the classical equations derived in the previous section, the quantum noise associated with the FWM process is captured \cite{Gordon_1963}. When the field ensemble is subject to loss through the fiber, fluctuations must be added to each quadrature of all field elements to ensure that the mean field power of the field ensemble converge toward the value of $\hbar \omega B_0/2$, the vacuum energy explicitly included in the ensemble, instead of zero. The loss fluctuation $\delta a_{\rm loss}$ follows a Gaussian distribution with the statistical properties \cite{Friis_2013}
\begin{eqnarray} \label{eq_35}
\langle \delta a_{\rm loss} \rangle &=& 0,\\
\langle \delta a_{\rm loss}^2 \rangle &=& \hbar \omega B_0\left[1-\exp(-\alpha \Delta z)\right]/4 \approx \hbar \omega B_0 \alpha \Delta z/4, \label{eq_26}
\end{eqnarray}
where $\Delta z$ is an infinitesimal piece of fiber in which the attenuation takes place, e.g. the numerical step size. The variance of the loss fluctuation in Eq. \eqref{eq_26} ensures that the mean power of the ensemble that undergoes a large loss converges towards $\hbar \omega B_0/2$. SpRS is included by adding another fluctuation to each quadrature of all field elements in the ensemble. Because the rates of SpRS on the S and aS sides are unequal, different fluctuation terms must be assigned to them: on the S side of a wave component $n$ of power $P_n$, the fluctuation $\delta a_{\rm Raman,S}$ must be added to each quadrature of a field $m$ with the properties (calculated in the Appendix)
\begin{eqnarray}
\langle \delta a_{\rm Raman,S} \rangle &=& 0,\\
\langle \delta a_{\rm Raman,S}^2 \rangle &\approx & \left[g^{(nm)}_{\rm R} P_n \Delta z (n^{(nm)}_{\rm T}+1)/2 - g^{(nm)}_{R}P_n\Delta z/4\right]\hbar \omega_{m} B_0,
\end{eqnarray}
where $n^{(nm)}_{\rm T}$ and $g^{(nm)}_{\rm R}$ depend on the frequency shift as explained above and $\Delta z$ is assumed small. On the aS side of wave component $n$, the corresponding fluctuation, $\delta a_{\rm Raman,aS}$, has the properties
\begin{eqnarray}
\langle \delta a_{\rm Raman,aS} \rangle &=& 0,\\
\langle \delta a_{\rm Raman,aS}^2 \rangle &\approx & \left( |g^{(nm)}_{\rm R}|P_n \Delta z\, n^{(nm)}_{\rm T}/2 + |g^{(nm)}_{R}| P_n \Delta z/4\right) \hbar \omega_{m} B_0.
\end{eqnarray}

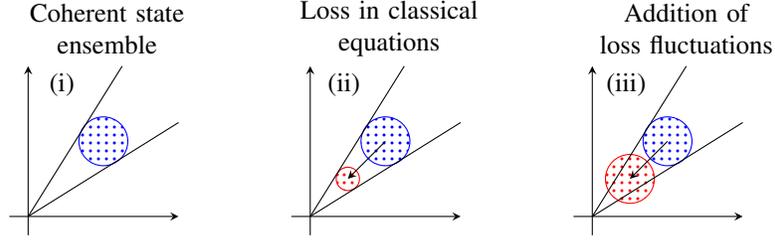
\begin{figure}[tb]
\centering
\begin{tikzpicture}[scale=1]

\begin{scope}[shift={(-3.75,0)}]
\draw [-stealth] (3.5,1.5)--(5.75,1.5);
\draw [-stealth] (3.75,1.25)--(3.75,3.5);

\draw [draw = blue,pattern = dots, pattern color = blue] (4.75,2.5) circle (0.325);
\draw (3.75,1.5)--(5.75,2.75);
\draw (3.75,1.5)--(5,3.5);

\node [text width = 2.5cm,align =center] at (4.8,4) {Coherent state ensemble};
\node at (4.2,3.25) {(i)};
\end{scope}

% UPPER MIDDLE PICTURE
\draw [-stealth] (3.5,1.5)--(5.75,1.5);
\draw [-stealth] (3.75,1.25)--(3.75,3.5);

\draw [draw = blue,pattern = dots, pattern color = blue] (4.75,2.5) circle (0.325);
\draw [draw = red,pattern = dots, pattern color = red] (4.25,2) circle (0.155);
\draw (3.75,1.5)--(5.75,2.75);
\draw (3.75,1.5)--(5,3.5);
\draw [-stealth,color=black] (4.75,2.5)--(4.25,2);
%\draw [stealth-] (4.5,2.15)--(5,1.75) node [right] {\small Attenuation};

\node [text width = 2.4cm,align = center] at (4.8,4) {Loss in classical equations};
\node at (4.2,3.25) {(ii)};

% UPPER RIGHT PICTURE
\begin{scope}[shift = {(-1,0)}]
\draw [-stealth] (8.25,1.5)--(10.5,1.5);
\draw [-stealth] (8.5,1.25)--(8.5,3.5);

\draw [draw = blue,pattern = dots, pattern color = blue] (9.5,2.5) circle (0.325);
\draw [draw = red,pattern = dots, pattern color = red] (9,2) circle (0.325);
\draw (8.5,1.5)--(10.5,2.75);
\draw (8.5,1.5)--(9.75,3.5);
\draw [-stealth,color=black] (9.5,2.5)--(9,2);

\node [text width = 2.5cm,align =center] at (9.75,4) {Addition of loss fluctuations};
\node at (8.95,3.25) {(iii)};
\end{scope}

\end{tikzpicture}
\caption{(i) A coherent state ensemble visualized in a phase-space diagram, (ii) how the ensemble is affected by loss in the classical equations, and (iii) the effect of adding loss fluctuations.} \label{fig_3}
\end{figure}

\noindent The effect of adding fluctuations during propagation is visualized for the loss process in Fig. \ref{fig_3}: in diagram (i), a coherent state ensemble of Eq. \eqref{eq_24} is visualized in phase space and in (ii), the ensemble is attenuated (blue$\rightarrow$red) through the classical Eq. \eqref{eq_105} in the absence of FWM and Raman scattering. Loss is classically a linear and phase-insensitive process, so all elements of the ensemble are translated directly towards the origin of phase space; this process is unphysical because the coherent state is squeezed in all directions simultaneously, thus breaking the uncertainty principle. In diagram (iii), the effect of adding the loss fluctuation of Eqs. \eqref{eq_35}--\eqref{eq_26} is seen to maintain the shape of the coherent state. The addition of the fluctuation ensures at the same time automatically that a NF equal to the loss is induced as expected of a passive device.

The ensemble approach described here has the advantage that it includes both amplitude and phase noise from SpRS in the FWM process. Raman amplifiers are usually described in power or photon number equations and the noise properties are derived from a statistical approach to the photon number mean and variance, which thus excludes phase noise.

%%%%%%%%%%%%%%%%%%%%%%%%%%%
%
%
% SEC 3: FIBER ATTENUATION
%
%
%%%%%%%%%%%%%%%%%%%%%%%%%%%
\section{Fiber attenuation}
In the context of realizing quantum state preserving frequency conversion, it is of interest to investigate how fiber loss affects the CE and NF of BS and how the loss fluctuations are coupled among the interacting waves in the FWM process, so we exclude Raman scattering and assume $\kappa = 0$ by one of the approaches discussed above.

\begin{figure}[!t]
\centering \begin{subfigure}{0.49\linewidth}  \centering
\includegraphics[width = \linewidth]{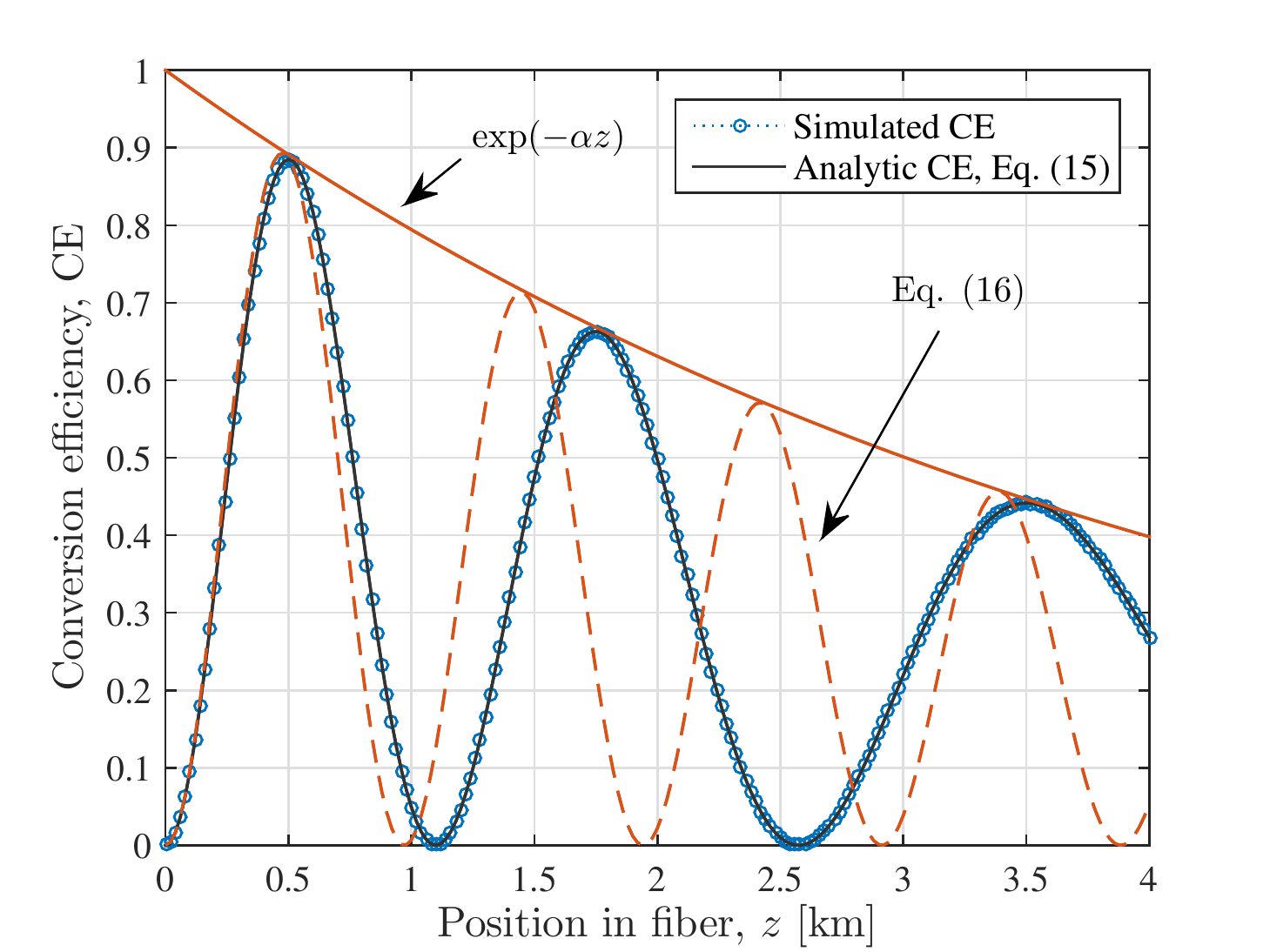}
\caption{}
\end{subfigure} %\hspace{0.1cm}
\begin{subfigure}{0.49\linewidth}  \centering
\includegraphics[width = \linewidth]{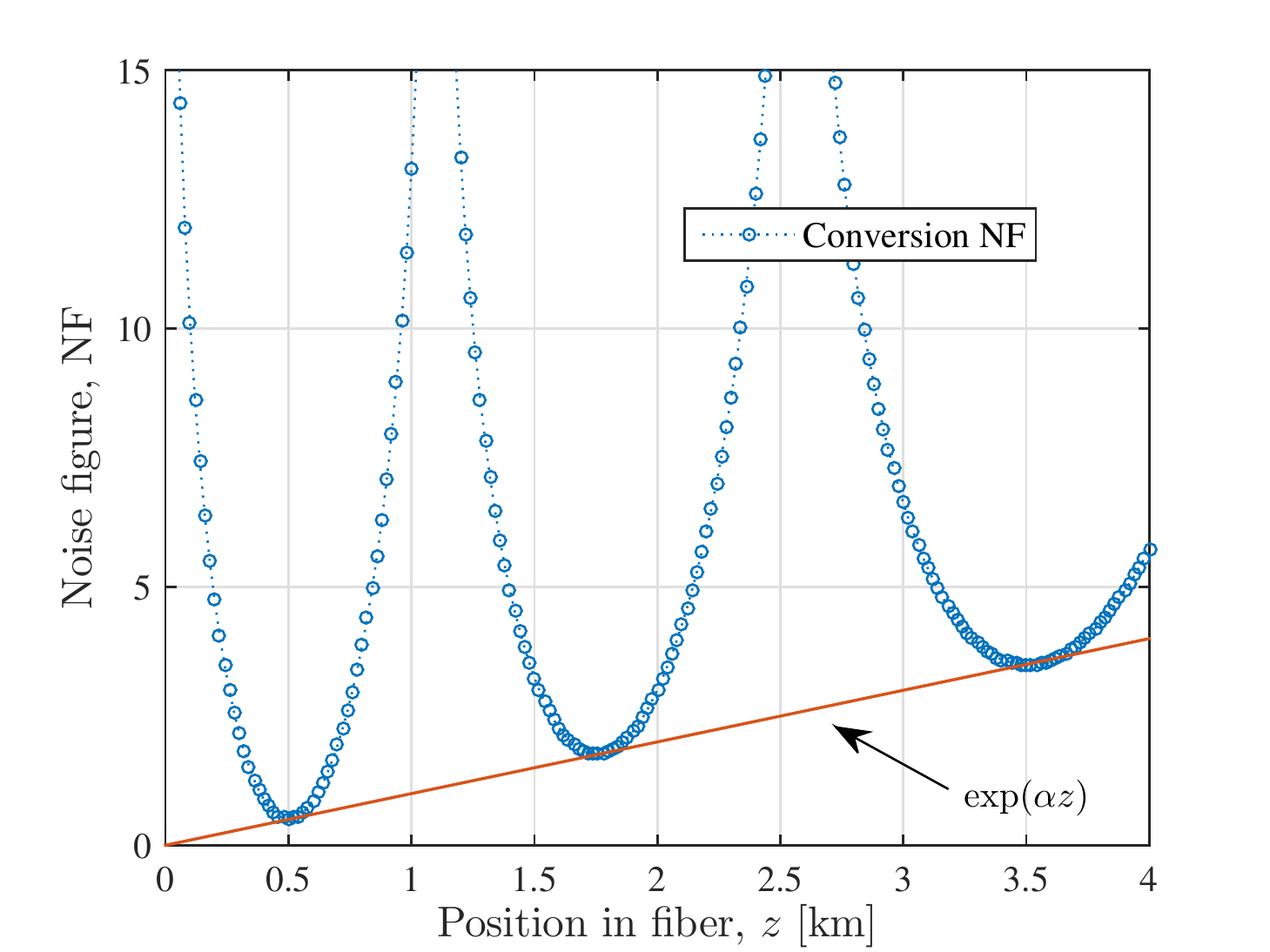}
\caption{}
\end{subfigure}\vspace{-0.25cm}
\caption{(i) CE versus fiber length with optimal phase matching and without Raman scattering. The red lines visually illustrate the effects of attenuation on the CE. (ii) The same for the conversion NF. In the simulation, $\alpha = 1$ dB/km, $P_{\rm p}=P_{\rm q} = 0.2$ W, $\gamma_{\rm s} = 9.89\ ({\rm W\;km})^{-1}$, $\beta_4 = 0$ ps$^4$/km, $\delta/2\pi = \Delta/2\pi= 1$ THz, and $\Delta z = 20$ m.} \label{fig_4}
\end{figure}

Figure \ref{fig_4} shows the results of solving Eq. \eqref{eq_105} (and the three corresponding ones of $\omega_2$--$\omega_4$) through a fiber of length $L = 4$ km and the addition of fluctuations during propagation with $\alpha = 1$ dB/km for all wave components, which implies that the results do not depend on which of the cases (a)--(d) is considered. The results depend only slightly on the values of $\delta$ and $\Delta$ through the explicit frequency dependence on the right hand side of Eq. \eqref{eq_105}. This frequency dependence is so small in highly nonlinear fibers that it is usually neglected \cite{Agrawal}. The CE versus position in the fiber is plotted in Fig. \ref{fig_4}(i) (blue dots) together with the analytic result of Eq. \eqref{eq_14} (solid black), where an excellent agreement is found. The red-dashed line compares the simulated CE to the approximation of Eq. \eqref{eq_13}, which gives the same as if losses on the pumps are neglected entirely; for small $\alpha z$ the approximation is seen to be reasonable. The importance of the solid-red line, the loss factor, is seen in Fig. \ref{fig_4}(ii) in which the loss-induced NF is plotted: the blue dots show the simulated NF versus position in the fiber where each local minimum corresponds to the maximum CE in Fig. \ref{fig_4}(i). The solid-red line thus marks a loss-induced noise floor that equals $\exp(\alpha z)$ and cannot be overcome. However, increasing the pump powers gives a shorter conversion distance since the conversion coefficient is $\mu^2 = \eta_{\rm i} \eta_{\rm s} \propto P_{\rm p}P_{\rm q}$, thus the accumulated signal and idler losses become smaller.

To investigate how the loss-induced noise couples between the signal, the idler, and the two pumps, we set the loss coefficient to zero at the different components in turn and repeat the simulation above (not shown graphically). If the losses of the signal and idler are excluded while keeping the losses of both pumps, the CE oscillates between 0 and 1 as if no loss was present, but the oscillation is still slowed by the loss in the pumps. The NF follows the CE, thus oscillating between infinity and 0. Hence, the loss-induced fluctuations in the pumps do not couple to the signal and idler. Turning on the loss in either of the signal or idler gives indistinguishable results, the characteristics of which are intermediate between turning the losses on or off in both. Consequently, the loss-induced fluctuations in the signal couple to the idler. If only the losses of the pumps are excluded, the slowing of the CE and NF oscillation disappears and the solution of Eq. \eqref{eq_13} is very accurate, but the NF observed in Fig. \ref{fig_4}(ii) is not decreased, which confirms that no pump fluctuations were coupled to the signal and idler through FWM.

%%%%%%%%%%%%%%%%%%%%%%%%%%%
%
%
% SEC 4: Stim. and spont. Raman Scat.
%
%
%%%%%%%%%%%%%%%%%%%%%%%%%%%
\section{Stimulated and spontaneous Raman scattering}
FWM and Raman scattering have a complicated interplay because they depend differently on the frequency shifts $\Delta$ and $\delta$: FWM requires phase matching, while SRS, which is anti-symmetric around the pumps, has a complicated material response in the frequency domain. Furthermore, as shown in Sec. \ref{sec_1}, SpRS is asymmetric around the pumps. To isolate the effects of SRS and SpRS in the FWM process, we disregard loss henceforth and assume again $\kappa = 0$.

\begin{figure}[!t]
\centering \begin{subfigure}[i]{0.49\linewidth}  \centering
\includegraphics[width = 0.99\textwidth]{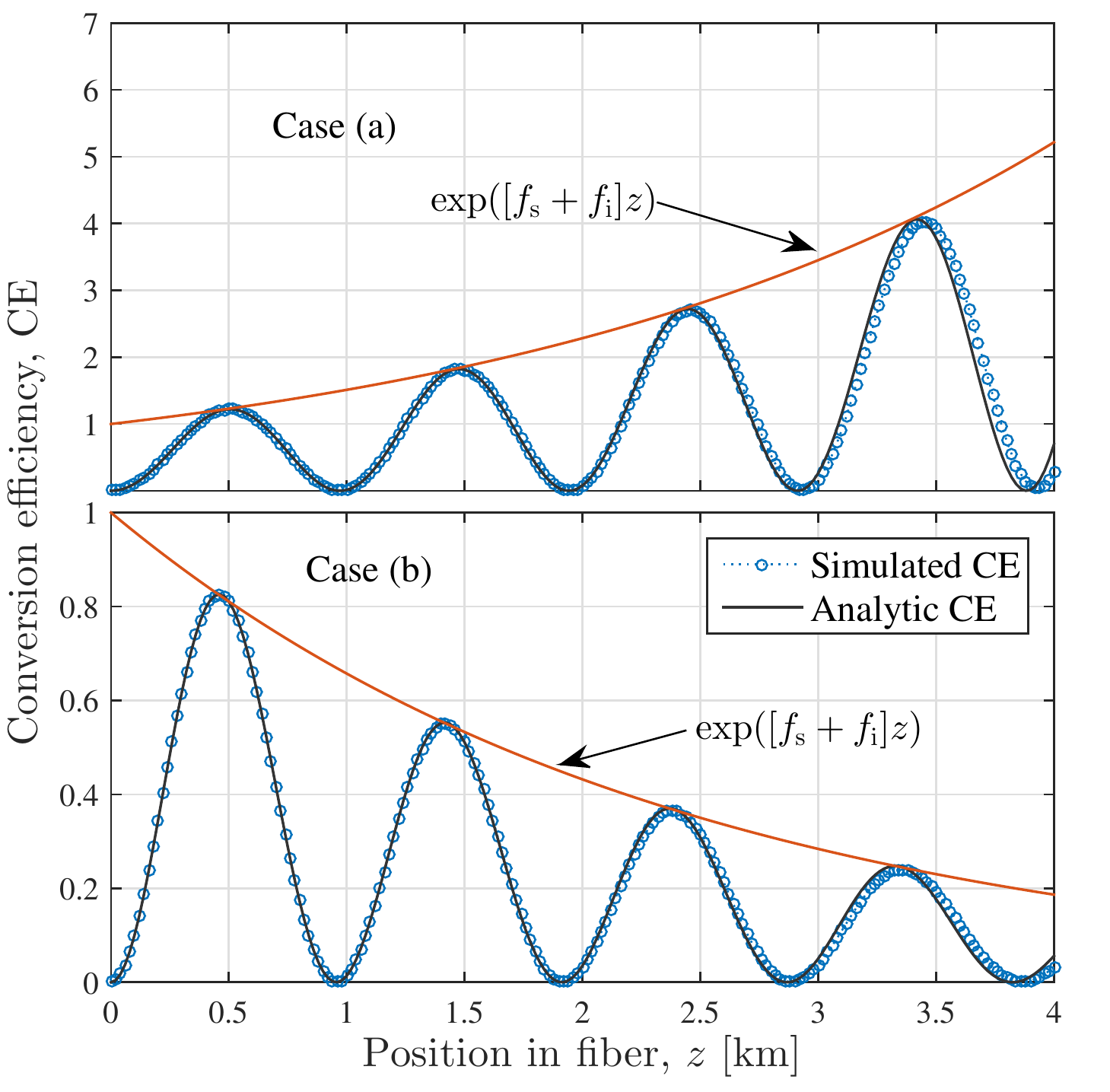}
\caption{ }
\end{subfigure} \hspace{\stretch{1}}
\begin{subfigure}{0.49\linewidth}  \centering
\includegraphics[width = 0.99\textwidth]{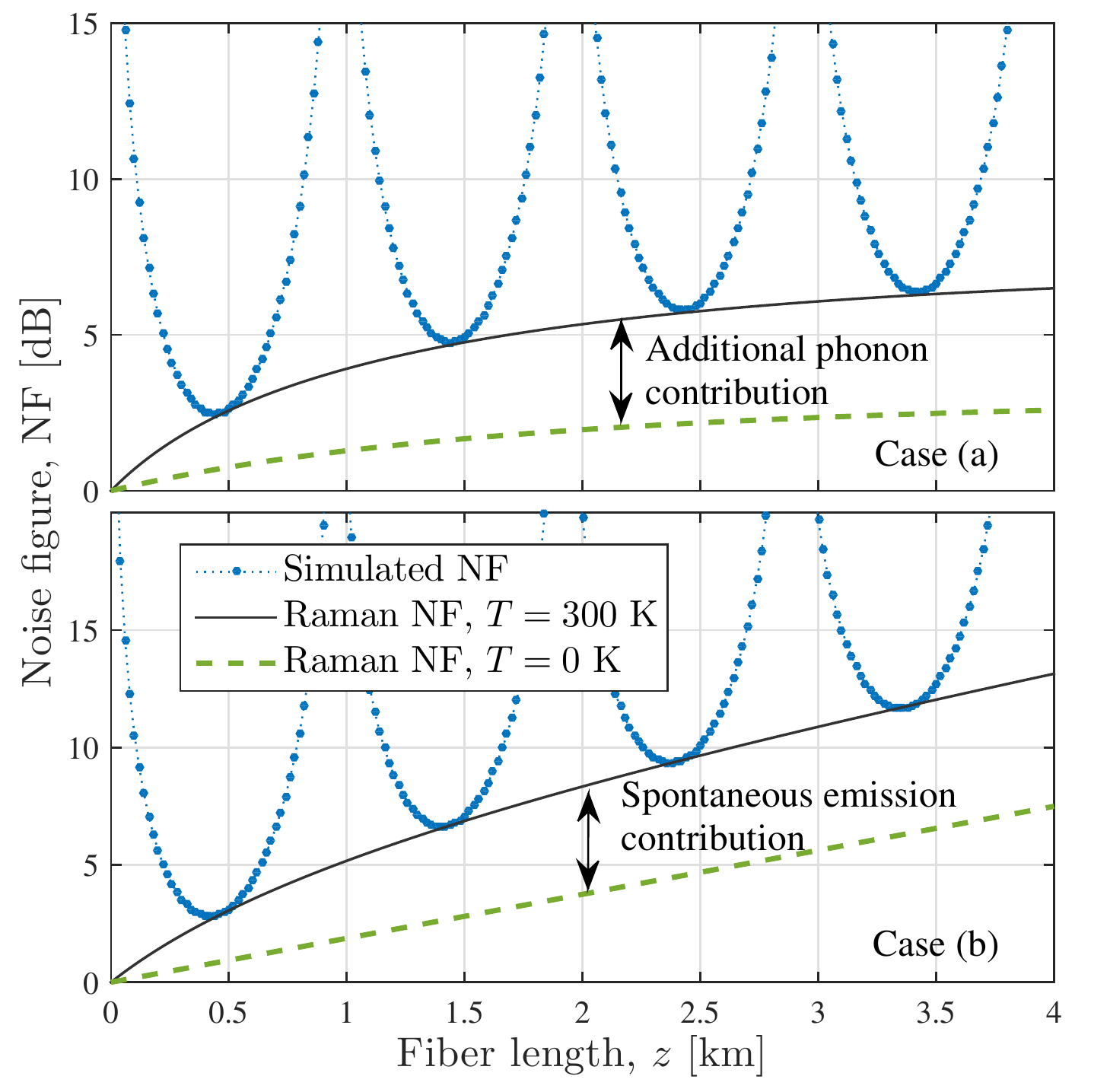}
\caption{ }
\end{subfigure}\vspace{-0.25cm}
\caption{(i) CE versus fiber length for case (a) (top) and (b) (bottom); both simulation (dotted blue) and analytic result of Eq. \eqref{eq_14} are shown; the legend applies to both plots, and the thick red line is the Raman amplification term. (ii) The same for the NF; the analytic Raman NF (solid black) is Eq. \eqref{eq_32} and Eq. \eqref{eq_33} for the top and bottom plots, respectively; the dashed green line is the Raman NF at 0 K. The parameters are the same as in Fig. \ref{fig_4}, but $\alpha= 0$ dB/km and $T = 300$ K.} 
\label{fig_5}
\end{figure}

Figure \ref{fig_5} shows the results of solving Eq. \eqref{eq_105} (and the three corresponding ones of $\omega_2$--$\omega_4$) through a fiber of length $L = 4$ km and adding Raman fluctuations at $T = 300$ K during propagation in cases (a) (top) and (b) (bottom); Fig. \ref{fig_5}(i) shows the CE versus position in the fiber, and the simulated (dotted blue) and the analytic (solid black) results agree very well. After $\sim 2$ km the two curves start differing significantly because it was assumed in the analytic expression that the pumps do not exchange energy. The actual energy exchange between the pumps, which is caused by Raman scattering, taking place in the simulation has two impacts on the conversion efficiency due to SRS transferring energy to the lower frequency pump, which is p in these cases: firstly, the difference in pump power causes a phase mismatch cf. the definition of $\kappa$; secondly, the conversion coefficient $\mu^2 \approx \eta_{\rm i} \eta_{\rm i} \propto P_{\rm p}P_{\rm q}$ becomes smaller because the product of two functions that have a constant sum is largest when values of the functions are equal.

In case (a), where the signal and idler are on the S side of the pumps, the CE grows through the fiber, essentially due to Raman amplification. That is, a CE higher than unity means that the signal and idler have been amplified by SRS such that the output idler has a higher output power than the signal input. The red curve denotes the Raman amplification term. The NF oscillates in phase with the CE, but the lowest achievable point increases according to the S side Raman-induced NF (calculated in the Appendix), valid for case (a),
\begin{eqnarray} \label{eq_32}
{\rm NF_{\rm S}} = \frac{1}{G}+ \frac{2[G-1](n_{\rm T}(\Omega_{\rm pi})+1)}{G},
\end{eqnarray}
where $G = \exp[g_{\rm R}^{\rm pi}(P_{\rm p}+P_{\rm q})z]$ is the Raman gain.
%and the S weighted phonon equilibrium number is 
%\begin{eqnarray}
%\tilde n_{\rm T}^{\rm S} = \frac{1}{f_{\rm i}+f_{\rm s}}\sum_{j={\rm p,q}} \frac{P_j}{2} \left[ g_{\rm \, R}^{j{\rm i}}(n_{\rm T}^{j{\rm i}}+1) + g_{\rm \, R}^{j{\rm s}}(n_{\rm T}^{j{\rm s}}+1) \right].
%\end{eqnarray}
The dashed green line of Fig. \ref{fig_5}(ii) (top plot) marks the Raman-induced NF at 0 K, so the region between that curve and the solid black curve is the NF induced by the existence of thermally excited phonons. %However, even at 0 K, SpRS contaminates the frequency conversion process on the S side.

In case (b), where the signal and idler are on the aS side of the pumps, the CE drops off exponentially because energy flows towards the pumps through SRS. The NF on the aS side is different from that on the S side for two reasons: firstly, since the aS process requires the presence of phonons, aS SpRS cannot occur at 0 K (as discussed above), and secondly, Raman depletion removes photons from a wave component so the SNR must change accordingly (much like the effect of losses). Hence, the aS Raman-induced NF (see the Appendix), valid for case (b), is
\begin{eqnarray}\label{eq_33}
{\rm NF_{aS}} &=& \frac{1}{D}+ \frac{2[1-D] n_{\rm T}(\Omega_{\rm ip})}{D},
\end{eqnarray}
where $D = \exp[-g_{\rm R}^{\rm ip}(P_{\rm p}+P_{\rm q})z]$. The first term on the right hand side, the Raman depletion term, is in contrast to the first term in Eq. \eqref{eq_32} growing through the fiber. The dashed green line of Fig. \ref{fig_5}(ii) (bottom plot) represents the minimum Raman NF of the aS side as caused by Raman depletion, and the region between that curve and the analytic curve is the entire SpRS contribution.

\begin{figure}[!t]
\centering 
\begin{subfigure}[i]{0.49\linewidth}  \centering
\includegraphics[width = 0.99\textwidth]{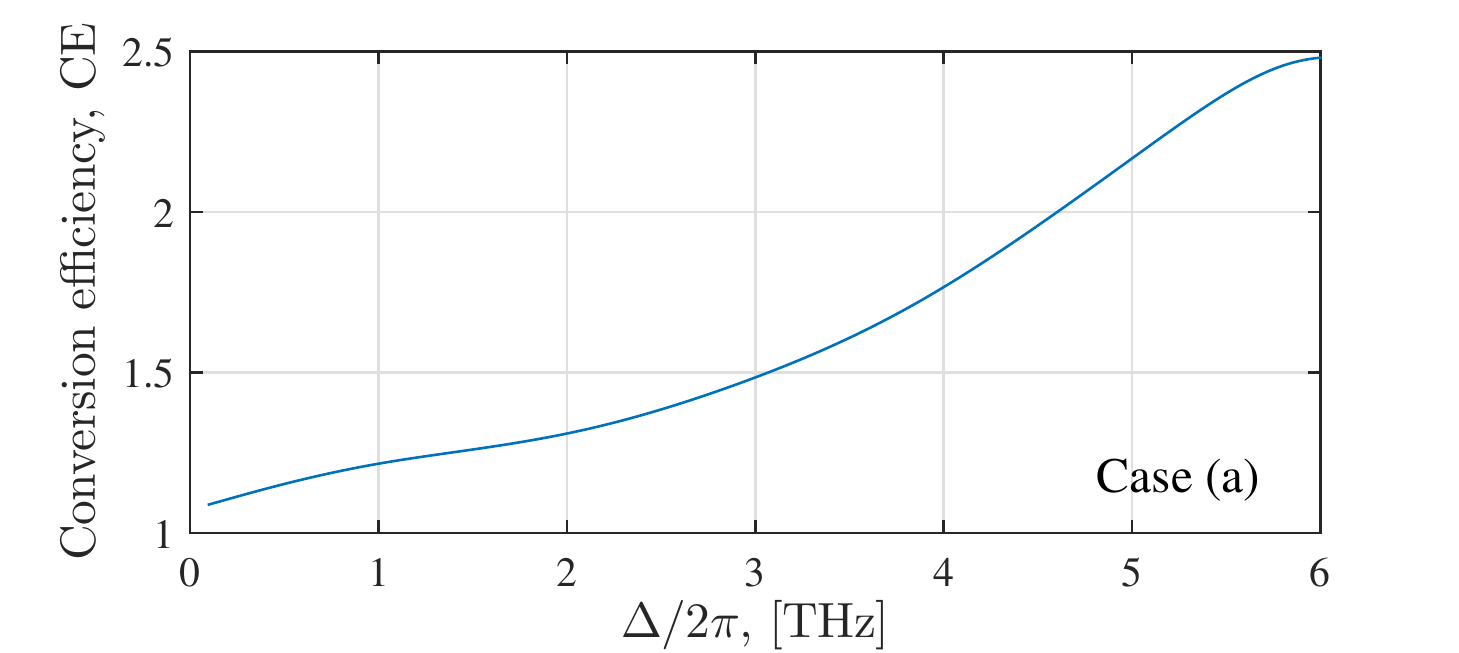}
\caption{ }
\end{subfigure} \hspace{\stretch{1}}
\begin{subfigure}{0.49\linewidth}  \centering
\includegraphics[width = 0.99\textwidth]{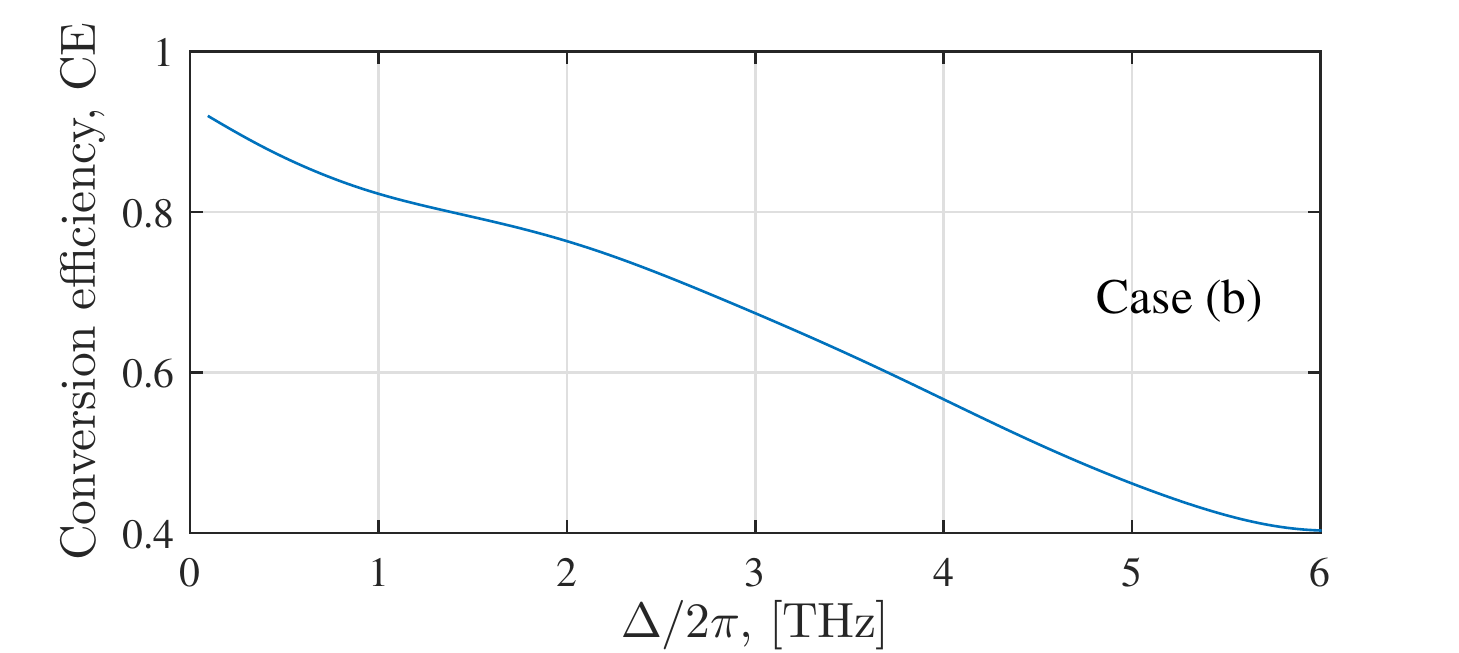}
\caption{ }
\end{subfigure}

\begin{subfigure}[i]{0.49\linewidth}  \centering
\includegraphics[width = 0.99\textwidth]{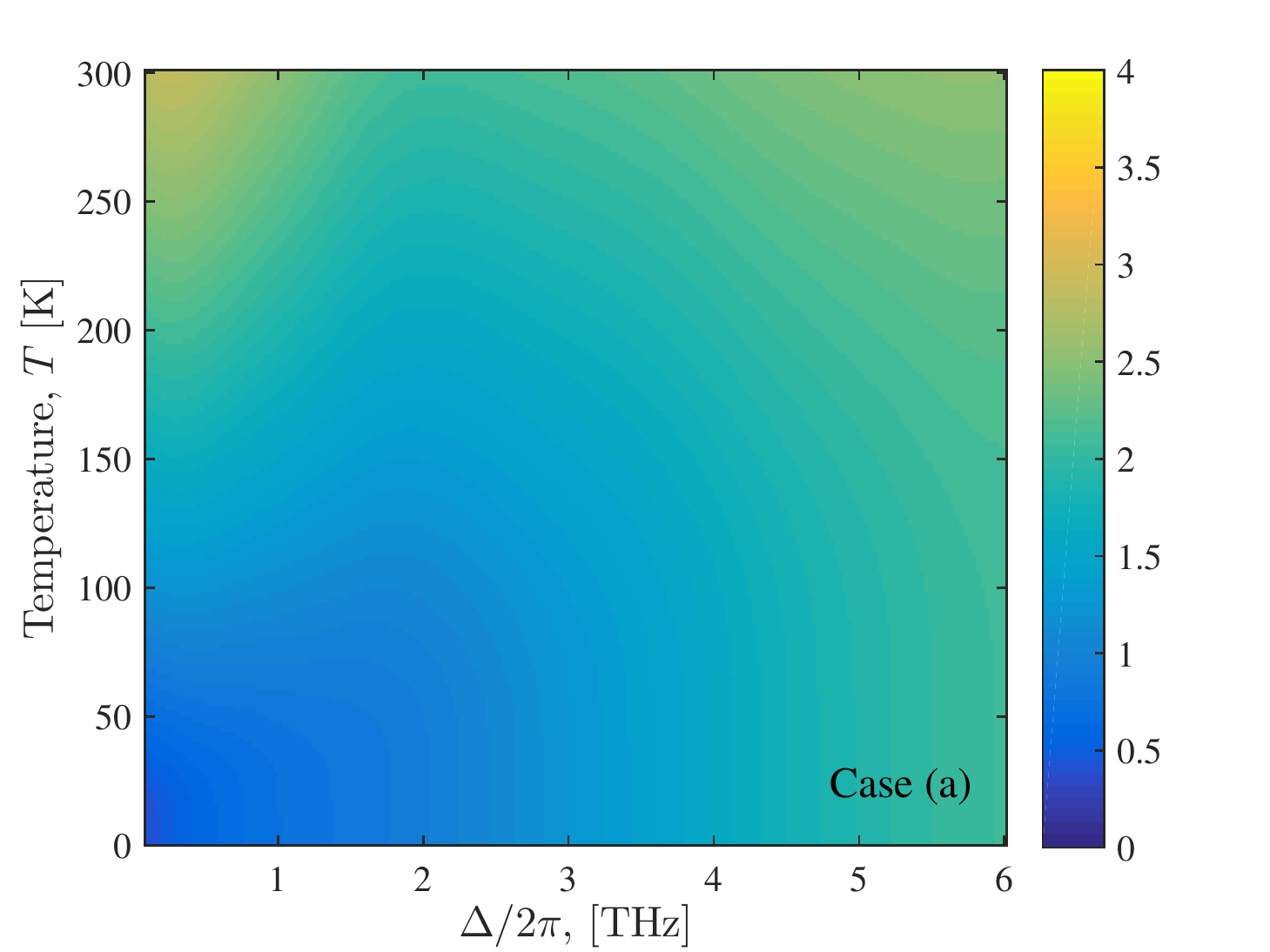}
\caption{ }
\end{subfigure} \hspace{\stretch{1}}
\begin{subfigure}{0.49\linewidth}  \centering
\includegraphics[width = 0.99\textwidth]{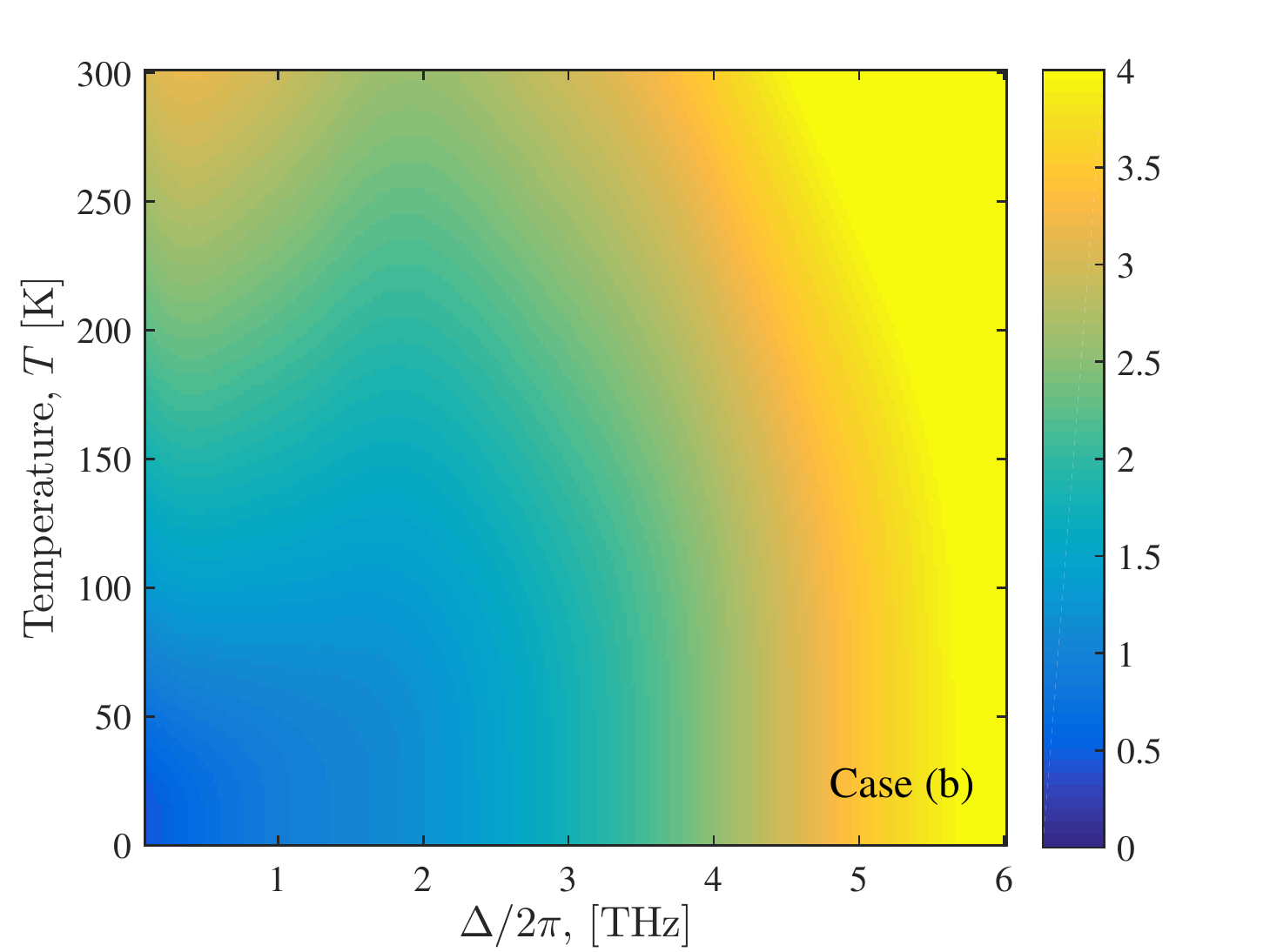}
\caption{ }
\end{subfigure}%\vspace{-0.25cm}
\caption{(i) and (ii) CE of Eq. \eqref{eq_14} of cases (a) and (b) of Fig. \ref{fig_1}, respectively; (iii) and (iv) NFs of Eqs. \eqref{eq_32} and \eqref{eq_33}, respectively, color scales are in dB.} 
\label{fig_6}
\end{figure}

Having established that Eq. \eqref{eq_14} is a good description of phase matched BS in the presence of Raman scattering and that the S and aS NFs of Eqs. \eqref{eq_32} and \eqref{eq_33} accurately predicts the NF of BS at the points of optimal conversion, we can use these analytic expressions to analyze the frequency and temperature dependencies of the BS CE and NF. Figure \ref{fig_6}(i) and (ii) shows the CE of Eq. \eqref{eq_14} versus $\Delta$ of cases (a) and (b) of Fig. \ref{fig_1}, respectively, at the fiber length of the first optimal conversion point, $L = \pi/(2\mu)$; $\kappa$ was assumed to be zero and \mbox{$P_{\rm p}=P_{\rm q} = 0.2$ W}, \mbox{$\delta = 1$ THz}, $\alpha = 0$ dB/km, and $\gamma_{\rm s} = 9.89 \ {\rm (W\, m)^{-1}}$ was used. The CE is in both cases essentially showing the signal and idler average Raman amplification, case (a), and depletion, case (b), that they receive from the two pumps; for these realistic parameter values, stimulated Raman scattering is a significant effect (in both cases up to a factor $\times 2.5$) that must be taken into account in silica fibers. Note that the frequency separation between the pump and the side band that are closest together is $2\Delta$.

Figure \ref{fig_6}(iii) and (iv) show the NF of cases (a) and (b), respectively, i.e. Eqs. \eqref{eq_32} and \eqref{eq_33} versus $\Delta$ and the temperature, $T$; the color scalings on the two plots are equal. For small $\Delta/2\pi$-values $<3$ THz, the two NFs are similar for all values of $T$, which is expected from Fig. \ref{fig_2} since the magnitudes of the rates of SpRS on the S and aS sides are comparable close to the pump; the aS side SpRS contribution is smaller but Raman depletion also contribute to the aS NF. The NF-values at low temperatures for small $\Delta$ (lower left corner) are in both cases approximately 0.4 dB; for small $\Delta$-values at room temperature (upper left corner) the NF is approximately 2.7 dB on the S side but 3.1 dB on the aS side. For larger $\Delta$-values (right side), the Raman depletion term becomes dominant on the aS side so the NF grows beyond 4 dB for all value of $T$. On the S side, where the NF is not affected by any Raman depletion term, the NF increases with $\Delta$ as expected from Fig. \ref{fig_2}. Given these numbers, the S side seems advantageous with lower NF but an important note on the difference in origin between the S and aS NFs should be taken. The S NF is solely induced by SpRS, which increases the power variance of a signal ensemble due to the random phase of the spontaneous decay; the aS NF is composed of both SpRS and the effect of Raman depletion. The rate of SpRS is always smaller on the aS side compared to the S side so the power variance increases less there; in the total NF of 3.1 dB at 300 K in \mbox{Fig. \ref{fig_6}(iv)}, SpRS is only responsible for 45\% of the total NF in linear units. This number is found by calculating the ratio of the temperature dependent part of Eq. \eqref{eq_33} (the second term) relative to the total $\rm NF_{aS}$.

At room temperature, the minimum achievable S and aS NFs are 2.1 dB and 2.6 dB, respectively, which underlines that Raman scattering has a significant Bragg scattering in terms of both CE and NF. Specific, realistic parameters where chosen in all simulations conducted in this paper but it has been verified in subsequent simulations that the presented results do not change notably by changing the $\gamma_j$ and the pump powers by an order of magnitude. Changing $\delta$ significantly leads to a different Raman interaction, especially between the pump and the side bands that are separated the most.

%%%%%%%%%%%%%%%%%%%%%%%%%%%
%
%
% Sec: Conclusion
%
%
%%%%%%%%%%%%%%%%%%%%%%%%%%%
\section{Conclusion}
We presented a model for calculating the quantum noise properties of parametric frequency conversion in form of Bragg Scattering; dispersion, loss, and stimulated and spontaneous Raman scattering were included. Closed-form analytic expressions for the conversion efficiency and Raman noise contribution were derived.

It was shown that loss in the signal and idler reduces the conversion efficiency and induces a noise floor equal to the common loss factor, and that loss in the pumps reduces the rate at which the energy oscillates between the signal and idler. Further elaboration showed that the pump losses do not induce a noise figure in the signal and idler. On the other hand, loss in either of the signal and idler induces a noise figure in the other component.

Both stimulated and spontaneous Raman scattering were shown to have significant impacts on parametric frequency conversion for a choice of realistic parameters for a highly nonlinear fiber: stimulated Raman scattering affects the conversion efficiency as would be expected from Raman amplifiers, the longer wavelength components receive energy from the shorter wavelength components. Spontaneous Raman scattering, which is asymmetric around the pumps, induces a minimum noise figure of 2.1 dB on the Stokes side of the pumps, case (a), for one full conversion from signal to idler at room temperature. On the anti-Stokes sides of the pumps, case (b), a minimum noise figure of $2.6$ dB caused by both spontaneous Raman scattering and Raman depletion is predicted. Lowering the temperature reduces SpRS on the Stokes side of the pumps but does not remove it; on the anti-Stokes side, SpRS is removed completely when lowering the temperature but the effect of Raman depletion is still present. 

Our theoretical predictions confirm that the presence of Raman scattering in silica fiber-based four-wave mixing in form of Bragg scattering contaminates the quantum noise-less frequency conversion to a significant degree that is comparable to the 3-dB noise figure induced by linear amplifiers such as phase-insensitive parametric amplifiers, Raman and Erbium-doped fiber amplifiers.

\section*{Acknowledgments}
The authors thank Colin J. McKinstrie for valuable discussions.

%%%%%%%%%%%%%%%%%%%%%%%%%%%
%
%
% Appendix
%
%
%%%%%%%%%%%%%%%%%%%%%%%%%%%
\appendix

\section*{Appendix: Variance of the Raman fluctuations and the Raman noise figure}
Given an ensemble of electromagnetic fields defined as in Eq. \eqref{eq_24}, we may apply a constant Raman gain $G$ in power from one pump to one signal, which leaves the SNR unchanged (signal and noise are amplified equally). The pump also provide spontaneous emission at the frequency of the signal ensemble, and this contribution changes the SNR. In this semi-classical model, we add fluctuation variables after having applied the gain and then calculate the corresponding change in the SNR. The output field ensemble of a Raman amplifier in the linear gain regime is
\begin{eqnarray} \label{eq_34}
A_{\rm out} = G^{1/2}\left[x_0 + \delta x+i(p_0+\delta p)\right] + \delta a_1 +i\delta a_2,
\end{eqnarray}
where $\delta a_j$ is the fluctuation in quadrature $j$ of the spontaneous emission associated with the physical process that provided the gain. Defining the SNR as Eq. \eqref{eq_36} and the Raman NF as the ratio $\rm SNR_{in}/SNR_{out}$, we get
\begin{eqnarray} \label{eq_25}
{\rm NF} = {\rm \frac{SNR_{in}}{SNR_{out}}} \approx 1+\frac{4\langle \delta a^2 \rangle}{G\, \hbar \omega B_0},
\end{eqnarray}
where a large photon number was assumed and it was furthermore assumed that the statistics of the fluctuations in both quadratures are equal, so $\langle \delta a_1^2 \rangle = \langle \delta a_2^2 \rangle = \langle \delta a^2 \rangle$. To evaluate the NF the variance of $\delta a$, which is equal to the second order moment  $\langle \delta a^2 \rangle$ because $\delta a$ has zero mean value, must be determined. 

Classical photon-number equations describing the Raman interaction between two waves at different wavelengths are shown in textbooks on nonlinear optics \cite{Rottwitt2015_bog} and it is custom to approximate the shorter wavelength component (the anti-Stokes component or pump in the context of Raman amplifiers) by a constant but here we also need to consider the opposite case of the longer wavelength component (the Stokes component) being much stronger than the shorter wavelength component and hence approximate that by a constant. Likewise, the mean output power associated with Eq. \eqref{eq_34} is
\begin{align} \label{eq_29}
\langle |A_{\rm S, out}|^2\rangle &= G(x_0^2 + p_0^2) + \hbar \omega_{\rm S}B_0 G/2 + 2\langle \delta a_{\rm S}^2 \rangle  \\
\langle |A_{\rm aS, out}|^2\rangle &= D(x_0^2 + p_0^2) + \hbar \omega_{\rm aS}B_0 D/2 + 2\langle \delta a_{\rm aS}^2 \rangle, \label{eq_30} 
\end{align}
where Eq. \eqref{eq_29} describes the weak Stokes component that receives Raman gain $G$ from the strong anti-Stokes component, and Eq. \eqref{eq_30} describes the distinct case of the weak anti-Stokes component that depletes by a factor $D$ by giving energy to the strong Stokes component. The second term of each equation is the amplification/depletion of the vacuum fluctuations explicitly included in Eq. \eqref{eq_34} and they are artifacts of the semi-classical modeling; $\langle \delta a_{\rm S}^2 \rangle$ and $\langle \delta a_{\rm aS}^2 \rangle$ should be chosen to counter-balance these terms as well to include spontaneous emission. If $\langle \delta a_{\rm S}^2 \rangle$ and $\langle \delta a_{\rm aS}^2 \rangle$ are chosen to be
\begin{eqnarray}
\langle \delta a_{\rm S}^2 \rangle &=& \left([ G-1](n_{\rm T}+1)/2 - [G-1]/4\right) \hbar \omega_{\rm S}B_0\\
\langle \delta a_{\rm aS}^2 \rangle &=& \left([1-D]n_{\rm T}/2 + [1-D]/4\right) \hbar \omega_{\rm aS}B_0,
\end{eqnarray} 
and they are inserted into Eqs. \eqref{eq_29} and \eqref{eq_30}, then one gets the equivalent of what may be derived from the classical equations. The Raman NF of a signal on either side of a strong pump is thus easily calculated using Eq. \eqref{eq_25} to be
\begin{eqnarray} \label{eq_38}
{ \rm NF_{\rm S}} &=& \frac{1}{G}+ \frac{2[G-1](n_{\rm T}+1)}{G} \rightarrow 2(n_{\rm T}+1)\\
{\rm NF_{\rm aS}} &=& \frac{1}{D}+ \frac{2[1-D]n_{\rm T}}{D}\rightarrow (1+2n_{\rm T})/D. \label{eq_39}
\end{eqnarray}
The arrows indicate the limits in case of large Raman interaction, i.e. $G \gg 1$ and $D \ll 1$. In the limit of low temperature, $n_{\rm T} \approx 0$, these formulas give the expected results of a 3-dB NF on the S side and a NF equal to the depletion on the aS side.

Considering case (a) and case (b) of Fig. \ref{fig_1}, in which there are two pumps and two small signals, the results Eqs. \eqref{eq_38}--\eqref{eq_39} do not immediately apply. However, if one regards the two pumps as one wave component with power $P = P_{\rm p} + P_{\rm q}$ and the signal and idler as one small signal, one has an artificial two-component system that may be described by the results of this appendix. The frequency separation between the two components is $\Omega_{\rm pi} = \omega_{\rm p} - \omega_{\rm i}$ in case (a) and $\Omega_{\rm ip}$ in case (b).
\end{document}